\documentclass[sn-mathphys-num]{sn-jnl}

\usepackage{graphicx}%
\usepackage{multirow}%
\usepackage{amsmath,amssymb,amsfonts}%
\usepackage{amsthm}%
\usepackage{mathrsfs}%
\usepackage[title]{appendix}%
\usepackage{xcolor}%
\usepackage{textcomp}%
\usepackage{manyfoot}%
\usepackage{booktabs}%
\usepackage{algorithm}%
\usepackage{algorithmicx}%
\usepackage{algpseudocode}%
\usepackage{listings}%
\usepackage{enumitem}

\theoremstyle{thmstyleone}

\theoremstyle{thmstyletwo}%

\theoremstyle{thmstylethree}%

\begin{document}

\title[Article Title]{Opportunity Cost in Insurance}

\author[1]{\fnm{Dr. Jan} \sur{Maelger}}\email{jan.maelger@t-online.de}

\affil[1]{Unaffiliated}
 
\abstract{We develop a formalism for insurance profit optimisation for the in-force business constraint by regulatory and risk policy related requirements. This approach is applicable to Life, P\&C and Reinsurance businesses and applies in all regulatory frameworks with a solvency requirement defined in the form of a solvency ratio, notably Solvency II and the Swiss Solvency Test. We identify the optimal asset allocation for profit maximisation within a pre-defined risk appetite and deduce the annual opportunity cost faced by the insurance company.} 


\maketitle

\section{Introduction}\label{sec1}
The goal of the senior management of any company ought to be to maximize shareholder value, achieved by maximizing long-term profits. Insurance companies are a priori no exception, but they do come with two special features. The first of which is that insurance is a heavily regulated industry, imposing many financial limits and constraints. 
The second is the role of the actuary, sitting in the extended management circle, equiped with veto rights and lawfully mandated to ensure sufficient levels of prudency. From a purist actuarial perspective, a too much in prudency buffers does not exist, causing from the outset a conflict of interest with the Finance and Investment departments. Qualitatively, the higher the prudency buffers, the higher the opportunity costs in foregone P\&L profits. With quantifications difficult, compromisary discussions are often vague, and a commercially-accessible concept of opportunity cost would prove useful.  

As seen from the viewpoint of a single insurance contract, its lifecycle can be split into the writing phase and its run-off phase. Similarly the business view separates the Underwriting from the management of the in-force portfolio, which require separate skills and have separate success factors. While insurance companies need to master both to prevail, it is viable to consider their optimization individually. 
On the Underwriting side, profitability is mainly driven by Pricing, on the optimization of which many studies have been performed \cite{Bauer2025, HenrietKlimenko, Harrington2013}.  Contrarily, on the management of the in-force business the field is largely unexplored. 

There are papers in the literatue on asset allocation optimization under insurance specific constraints, but they differ from our study by not deriving a concept of opportunity cost and in either restricting to only solvency constraints or in optimizing for individual stock selections instead of the entire SAA. For example, in \cite{SchluetterFianuGruendl} the authors study an optimal stock portfolio selection under general solvency constraints and in the context of ESG-investing. In \cite{BraunSchmeiserSchreiber} the insurance asset allocation is optimized across all major asset classes under the constraint of adhering to Solvency II market risk capital requirements. 

The approach developed here is interesting for institutional investors, who have the means to run the analysis on a large scale gauging an insurance companies` potential foregone growth due to sub-optimally chosen metrics, or the insurance companies themselves, looking to gain further insight into either their own or a competitor`s strategy. 

Following the school of Carl Bender and Steven Orszag \cite{BenderOrszag} to prioritize approximative answers over exact non-answers, we abstain from rigorous definitions and a theorem-proof-style discussion. Instead, we focus on practical applicability and always assume sufficient smoothness. The more notions are kept vague, the more powerful the approach, and so this study is meant as a fall-back blue-print for practicioners of either insurance companies or institutional investors attempting to better understand the business at hand. 

In section \ref{sec2} we develop the formalism to compute the opportuntiy cost faced by an insurance company for generic profit functions and a generic set of constraints. In section \ref{sec3} we discuss various choices of profit functions and in section \ref{sec4} we give an overview of the most common constraint types. Section \ref{sec5} explains how time-dependence enters the equations and in section \ref{sec7} we comment on local approximations. Some concluding remarks are made in section \ref{sec13}.

\section{Generic Formalism}\label{sec2}
Let $P$ denote the profit of an insurance company in a given reporting cycle, typically a year or a quarter, in dependence of some vector of variables $\underline{x}$ and external parameters $\underline{\tau}$: $P(\underline{x}, \underline{\tau})$, subject to a set of constraints $c$ with $$c_i(\underline{x}, \underline{\tau}) = 0 \quad\text{and}\quad c_j(\underline{x}, \underline{\tau}) \geq h_j\Leftrightarrow \tilde c_j(\underline{x}, \underline{\tau})\equiv h_j-c_j(\underline{x}, \underline{\tau})\leq 0$$ for some indices $i,j \in \mathbb{N}$ and thresholds $h_j\in \mathbb{R}^+$. The constraints may be of regulatory nature, eg. on solvency or liquidity, or internally set limits within a risk appetite statement. A more detailed discussion of the constraints can be found in section \ref{sec4}. Denote by $n$ the total number of constraints.   

All quantities immediately controllable by the management of the company constiute the variables $\underline{x}$, in particular the strategic asset allocation (SAA), and all quantities outside the influence of the management give the set of external parameters $\underline{\tau}$, e.g. the yield curve, credit spreads or reserving assumption parameters. 
With the $\underline{x}$ being the asset class allocations, they range from 0 to 1, ie $\underline{x}\in [0;1]^{|x|}$, and they must sum to 1 for consistency, creating the additional constraint $$\sum_k x_k -1 =0.$$

Since we consider the situation of in-force business instead of underwriting new one, the contractual terms of the in-force policies are also seen as external paramters. In particular, we disregard Pricing effects on profit optimisation in this study and refer interested readers to a small selection of the vast literature available \cite{Bauer2025, HenrietKlimenko, Harrington2013}. Similarly we also regard process operations in the Claims department as given constants. While efficiency gains in claims handling obviously reduce costs and increase profits, such undertakings are typically multi-year projects and not immediate levers for the senior management to pull.  

Mathematically, the constraint profit optimisation problem can be formulated via Lagrange multipliers \cite{Beavis_Dobbs_1990,ProtterIntermediateCalc} under Karush-Kuhn-Tucker (KKT) conditions \cite{Karush2014MinimaOF,Tind2009,Lange2013, walsh1975methods}. We identify the insurance Lagrangian function as
\begin{equation}\label{eq1}
	\mathcal{L}(\underline{x},\underline{\tau},\underline{\lambda})= P(\underline{x}, \underline{\tau})-\sum_{i}\lambda_i\, c_i(\underline{x},\underline{\tau})-\sum_{j}\tilde\lambda_j\, \tilde c_j(\underline{x},\underline{\tau})-\mu \big(\sum_k x_k -1\big),  
\end{equation}
for some multipliers $\lambda_i,\tilde\lambda_j,\mu\in \mathbb{R}$ and 
$\underline{\lambda}=\begin{pmatrix}\lambda_i\\ \tilde\lambda_j\\ \mu\end{pmatrix}$.
By the KKT theorem, any saddle point\footnote{Maximum over $\underline{x}$, but could be a maximum or minimum over $\underline{\lambda}$.} of the Lagrangian over $\underline{x}\oplus\underline{\lambda}$ is also an optimal solution over $\underline{x}$ to the profit function subjet to the above constraints \cite{walsh1975methods,Kalman01062009,Fuente_2000}. 
Reformulated as a KKT problem, we have the following necessary conditions for an optimal solution \cite{Ruszczynski}: 
\begin{itemize}
	\item Stationarity condition:
	\begin{eqnarray}\label{e7}
		\nabla_{x}P(\underline{x}, \underline{\tau})
		-\sum_{i}\lambda_i\, \nabla_{x}c_i(\underline{x},\underline{\tau})-\sum_{j}\tilde\lambda_j\, \nabla_{x}\tilde c_j(\underline{x},\underline{\tau})-\mu = \underline{0}      
	\end{eqnarray}
	\item Primal feasability conditions (original constraints):
	\begin{eqnarray}\label{e8}
		\sum_k x_k -1=0\quad\text{and}\quad c_i(\underline{x}, \underline{\tau}) = 0 \quad\text{and}\quad  \tilde c_j(\underline{x}, \underline{\tau})\leq 0\quad\forall i,j\label{e9}
	\end{eqnarray}
	\item Dual feasability condition:
	\begin{equation}\label{e10}
		\tilde\lambda_j\geq 0 \quad \forall j 
	\end{equation}
	\item Complementary Slackness condition:
	\begin{eqnarray}\label{e11}
		\sum_{j}\tilde\lambda_j\,\tilde c_j(\underline{x},\underline{\tau}) = 0
	\end{eqnarray}
\end{itemize}
The notation $\nabla_x$ stands for a vector of partial derivatives over all elements of the vector $\underline{x}$.
The sufficient conditions for an optimal solution can be expressed via the determinant of the Hessian matrix \cite{Boyd_Vandenberghe_2004, silberberg2001structure} as
\begin{equation}\label{e12}
	\det \nabla^2_{xx} \mathcal{L}(\underline{x},\underline{\tau},\underline{\lambda}) < 0.  
\end{equation}
To give some intuition, the potential candidates (ie stationary points) for a global maximum of the Lagrangian in Eq.(\ref{eq1}) are found by the solutions of Eqs.(\ref{e7})-(\ref{e11}). These stationary points could correspond to either local maxima, minima or saddle points over $\underline{x}$. Eq.(\ref{e12}) then sorts out the local maxima from the pool of candidate solutions. To identify the global maximum $\underline{x}_\ast$, one has to manually compare all local maxima such that $\mathcal{L}(\underline{x}_\ast)\geq \mathcal{L}(\underline{x}_m)$ for all $\underline{x}_m$ satisfying Eqs.(\ref{e7}) to (\ref{e12}). In case there are no candidate solutions $\underline{x}_m$, then the global maximum $\underline{x}_\ast$ sits on the boundary of the variable space $[0;1]^{|x|}$. 

The difference bewtween the profits generated from the optimal target SAA values and the actuals is the opportunity cost faced by the insurance company: 
\begin{equation}\label{e13}
OpC(\underline{x},\underline{\tau})\equiv P(\underline{x}_\ast,\underline{\tau})-P(\underline{x},\underline{\tau}).\end{equation}   
It is remarked that the optimization problem in Eqs (\ref{e7}) to (\ref{e12}) is always well-defined for any chosen number of constraints $n_c$ and $n_{\tilde c}$ and for any chosen granularity\footnote{For example $\text{SAA}=\big\{\text{Cash, RE, Equities, Alternatives, Govies, Coprorate\, bonds}\big\}$.} of the SAA, with the number of classes denoted by $n_x$. The number of equations and the number of unknowns are both $n_c+n_{\tilde c}+n_x$.

In general, it is not possible to solve the set of Eqs. (\ref{e7}) to (\ref{e12}) analytically and one resorts to numerical methods. A vast array of techniques are available in the literature, and the reader is referred to \cite{Boyd_Vandenberghe_2004} for an introduction to nonlinear optimization and to \cite{MOHAMMADI2023106959,NeculaiAndrei} for a review of the more recent developments. Finally, the advent of commercially viable quantum computing may bring further capabilities to real-time financial optimization \cite{Abbasetal,WilkensMoorhouse}.

\section{Choosing a Profit Function}\label{sec3}
Insurance profits can be split in many ways, and accounting nomenclature can become highly involved. However, at lowest granularity, one way to determine profits is via the value of the new business written ($NBV$), the change in existing assets on the balance sheet from one period to the next, i.e. the investment returns ($InvR$), both realized and unrealized, and the change in liabilities during the same time period ($\Delta liab$): 
	\begin{equation}\label{e1}
	P=\textit{NBV}+\textit{InvR}-\Delta\textit{liab}.
\end{equation} 

In itself, each of these components is a complex function of many stochastic drivers. Eg. the InvR depend heavily on the SAA and the market developments. Similarly the movements of the liabilities depends on the market developments and the change in technical reserving assumptions and parameters. Additionally, via measures such as Asset Liability Management (ALM), the adverse changes in liabilities can be mitigated by mirroring changes in assets, adding a further layer of complexity onto the business. 

We disregard $NBV$ at this point, since it is primarily determined by Pricing and thus out of scope for this study. The change in liabilities is itself composed of several parts, most of which are outside the sphere of influence of the senior management, eg the change in reserving assumptions, with an exception of operational or one-off reserve releases. However, the former is simply balancing items and the latter can be seen as deferred profits from the past. The term $\Delta\textit{liab}$ also contains a contribution from the difference in actual claims incurred vs the reserved forecast from the previous reporting date. This is largely due to the random claims experience during the reporting period. However a small part is influencable by the senior management via efficient internal processes and good preventive communication to the customers, eg early hail warnings to motor customers. Nonetheless, in the spirit of this paper, these effects are attributed in its entirety to the external parameters $\underline{\tau}$. 

For the investment returns, they can be written as 
\begin{equation}\label{e14}
	\textit{InvR}=\sum_k^{|x|} R_k x_k
\end{equation}
with $R_k$ a random variable for the investment returns of asset class $x_k$ over a given time period, which can be expressed as $R_k=\mathbb{E}[R_k]+v_k$ with the volatility $v_k$ a stochastic noise around 0. At first order, one may set $v_k=0$, when considering long-term over short-term profits. Since the market returns are outside the influence of the management, the $R_k$ are part of the set of external parameters, $R_k\in\underline{\tau}$.

Looking at profit functions in more detail, there is a whole plethora of metrics available, for instance 
\begin{itemize}
	\item Business Operating Profit (BOP)
	\item Net income before interest and tax (NIBIT)
	\item Net income after tax (NIAT)
	\item Return on Equity (ROE)
	\item Return on Capital (ROC)
	\item Risk-adjusted return on capital (RAROC)
	\item Return on risk-adjusted capital (RORAC)
	\item Risk-adjusted return on risk-adjusted capital (RARORAC)
\end{itemize}
These profit metrics can be applied to balance sheets under different accounting standards, either statutory (eg. US GAAP, Swiss GAAP etc.) or market consistent (eg. IFRS, Swiss GAAP FER, SII etc.)  

All of these choices hold merit in their own right and will ultimately lead to different opportunity costs. For example, choosing to consider BOP on an IFRS balance sheet yields a respective opportunity cost 
$OpC_{\text{IFRS BOP}}$.

\section{Constraints}\label{sec4}
Insurance is a heavily regulated industry and the list of constraints imposed by the regulator is long. Many of the regulatory constraints are superimposed by internally set confidence (amber/red) limits, specified in risk appetite statements, to ensure sufficient early warning indicators before any regulatory breaches. 
We list below some of the most common limit types, without any claim to even near exhaustion. 

\begin{enumerate}
	\item Cash requirement: $$x_\text{Cash}\geq l_\text{Cash},$$
	with $x_\text{cash}$ the SAA component for cash and cash equivalents, and $l_\text{Cash}$ the short-term cash limit.
	
	\item Liquidity strain requirement: $$\sum_{j\in\text{liquid}} x_j h_j \geq l_\text{liquid}$$
	The quantity $l_\text{liquid}$ is the long-term lower liquidity threshold, and the $\big\{h_j|j\in{\text{liquid}}\big\}$ are haircuts on the market values of asset classes $\big\{x_j |j\in\text{liquid}\big\}\subset\underline{x}$, inserted to account for potential losses realized during the sale of liquid assets in stress situations. 
	
	\item Solvency requirement: $$SR(\underline{x})\equiv AFR(\underline{x})\big/SCR(\underline{x})\geq T$$
	The solvency constraint is in the usual form of a solvency ratio ($SR$) with $AFR$ are the available financial ressources, $SCR$ the solvency acapital requirement\footnote{The $AFR$ is called Basic Own Funds ($BOF$) under SII and Risk Bearing Capital ($RBC$) under SST. The SII wording of SCR corresponds to the Target Capital ($TC$) under SST. Conceptually, for our purposes, these terms are interchangeable.} and the parameter $T$ either the minimum regulatory threshold or an internally set target solvency ratio.
\end{enumerate}
Further typical constraints include thresholds for the following quantities.
\begin{enumerate}[resume]	
	\item Shareholder Equity
	\item Tied Assets
	\item FX Exposure
	\item Counterparty Exposure
	\item ALM or Hedging requirements
	\item Usage of derivative instruments
	\item Cash pooling limits
\end{enumerate}
While asset class exposure limits may already be in place in the company, such as $x_\text{Equities} \leq l_\text{Equities}$ or $x_\text{RE} \leq l_\text{RE}$ etc, one should be careful in modelling them as constraints in Eq.(\ref{eq1}), since they artificially restrict the optimizable variable space. If these asset class limits are rooted in regulatory requirements, they must of course be adhered to. Contrarily, if they merely stem from the current SAA, then this is in fact the setup to be challenged by Eqs.(\ref{e7})-(\ref{e12}) and their inclusion as constraints leads to a misinterpretation of the true maximum $x_\ast$.

\section{Time-dependence}\label{sec5}
Investigating the time-dependence of Eq.(\ref{eq1}), the only quantities which change with time are the external parameters, $\underline{\tau}(t)$. There is no explicit time-dependence of the profit function $P$ or the variables $\underline{x}$, such that $\frac{\partial P}{\partial t}=0$ and 
\begin{equation}\label{eq19}
	\frac{\text{d} P}{\text{d} t}=\sum_{k=1}^{|\tau|}\frac{\partial P}{\partial \tau_k}\frac{\text{d} \tau_k}{\text{d} t}.
\end{equation}

Similarly for the constraints $c_i(\underline{x}, \underline{\tau})$ and $\tilde c_j(\underline{x}, \underline{\tau})$.
In a strict sense, the time-dependence turns Eqs.(\ref{eq1}) to (\ref{e12}) into a real-life optimisation problem to be solved. However, while market data indeed changes in real-time, most of the external paramters, ie reserving assumptions or yield curves, are more stable, rendering quarterly updates sufficient in most scenarios.   

\section{Local Taylor expansion}\label{sec7}
By standard Taylor expansion, around any point $\underline{x}_0$, the $|x|$-dimensional hypersurface $\mathcal{L}(\underline{x})$ can be approximated by 
\begin{equation}\label{e14}
	\mathcal{L}(\underline{x},\underline{\tau})=\mathcal{L}(\underline{x}_0,\underline{\tau})+\sum_{k=1}^{\infty}\nabla_x^k\mathcal{L}(\underline{x}_0,\underline{\tau}) \cdot (\underline{x}-\underline{x}_0)^k,  
\end{equation}
which is valid for any $\underline{x}$ in sufficient proximity of $\underline{x}_0$.
With a closed form not known for $\mathcal{L}$, decision making processes between a handful of alternatives typically assess their impacts on the profits $P$, accompanied by impact estimates on other KPIs. Implicitly, such impact assessments correspond to a subset of the linear terms $\nabla_x\mathcal{L}(\underline{x}_0,\underline{\tau})$. In rare cases, second order effects are accounted for to enhance decision quality. 
While this approach is practical, it only gives direction for a small set of possible choices, without clear identification of true maxima. 
At any point $\underline{x}_0$, the best choice is the one that follows the gradient of the surface, $\nabla_x\mathcal{L}(\underline{x}_0,\underline{\tau})$.

\section{Conclusion}\label{sec13}

We have set up a universal formalism for any insurance company to tune its strategic asset allocation for profit maximization, or at least to better understand the opportunity costs associated with differing strategies. The equations in this conceptual first study remain at high-level, but can be tailored to fit specific business needs and with some ressource investment can be turned into regular reporting outputs.

While the formalism developed allows to identify the optimal strategic asset allocation for maximum profits within any set of constraining limit thresholds, it does not make any claim how to best set the limits themselves. This remains for a future study.    

\backmatter

\bmhead{Acknowledgements}

The author would like to thank James Richardson for mentorship and practical insights into the steering of an insurance company as seen from an actuarial background. This publication did not receive any financial support.





\bibliography{sn-bibliography}

\end{document}